\documentstyle[aps,prd,eqsecnum,amssymb,floats,epsfig]{revtex}

\newcommand{\be}{\begin{equation}}
\newcommand{\ee}{\end{equation}}
\newcommand{\bea}{\begin{eqnarray}}
\newcommand{\eea}{\end{eqnarray}}

\newcommand{\dst}{\displaystyle}
\newcommand{\pa}{\partial}
\newcommand{\ve}{\varepsilon}

\def\httiv#1#2{h^{\rm TT}_{(4)#1#2}}
\def\ghttiv#1#2#3{h^{\rm TT}_{(4)#1#2,#3}}
\def\dothttiv#1#2{\dot{h}^{\rm TT}_{(4)#1#2}}

\def\pitiii#1#2{\widetilde{\pi}^{#1#2}_{(3)}}

\def\siv{S_{(4)}}
\def\sij#1#2{S_{(4)#1#2}}
\def\gsij#1#2#3{S_{(4)#1#2,#3}}
\def\gsiv#1{S_{(4),#1}}
\def\fii{\phi_{(2)}}
\def\gfii#1{\phi_{(2),#1}}
\def\fiv{\phi_{(4)}}
\def\gfiv#1{\phi_{(4),#1}}

\def\v#1{V_{(3)}^#1}
\def\gv#1#2{V_{(3),#2}^#1}

\begin{document}

\title{
Dimensional regularization of the gravitational interaction of point masses}

\author{Thibault Damour$^{1)}$,
        Piotr Jaranowski$^{2)}$ and
        Gerhard Sch\"afer$^{3)}$}

\address{
{\sl $^{1)}$ Institut des Hautes \'Etudes Scientifiques,
91440 Bures-sur-Yvette, France}\\
{\sl $^{2)}$ Institute of Theoretical Physics, University of Bia\l ystok,
Lipowa 41, 15-424 Bia\l ystok, Poland}\\
{\sl $^{3)}$ Theoretisch-Physikalisches Institut, 
Friedrich-Schiller-Universit\"at, Max-Wien-Platz 1, 07743 Jena, Germany} }

\maketitle

\begin{abstract}

We show how to use dimensional regularization to determine, within the
Arnowitt-Deser-Misner canonical formalism, the reduced Hamiltonian describing
the dynamics of two gravitationally interacting point masses.  Implementing, at
the third post-Newtonian (3PN) accuracy, our procedure we find that dimensional
continuation yields a finite, unambiguous (no pole part) 3PN Hamiltonian which
uniquely determines the heretofore ambiguous ``static'' parameter:  namely,
$\omega_s=0$.  Our work also provides a remarkable check of the perturbative
consistency (compatibility with gauge symmetry) of dimensional continuation
through a direct calculation of the ``kinetic'' parameter $\omega_k$, giving the
unique answer compatible with global Poincar\'e invariance
($\omega_k=\frac{41}{24}$) by summing $\sim50$ different dimensionally continued
contributions.

\end{abstract}

\section{Introduction}\label{sec1}

The problem of the gravitational interaction, within Einstein's theory, of two
bodies has been revived by the realization that the most promising candidate
sources for ground-based interferometric gravitational-wave detectors such as
LIGO and VIRGO are binary black holes.  The most significant part of the
gravitational-wave signal emitted by two black holes corresponds to the last few
orbits before coalescence.  This makes it urgent to derive the equations of
motion of two black holes with the highest possible accuracy.  The equations of
motion of two compact bodies (black holes or neutron stars) at the 2.5
post-Newtonian (2.5PN) approximation were first obtained in Refs.\
\cite{DD81,D82,D83}. [We recall that ``$n$PN approximation'' refers to
the terms of fractional order $(v/c)^{2n}\sim\left(GM/(c^2r)\right)^n$ in the
equations of motion.]  Of special importance for the following was the proof
\cite{D83} that the property of ``effacement'' of the internal structures of
(non-spinning) gravitationally interacting compact bodies implied that their
equations of motions depended only on their two ``Schwarzschild'' masses $m_1$,
$m_2$ up to the 5PN level.  This effacement property was used to prove in
\cite{D83} that the 2.5PN equations of motion of two compact objects were
correctly obtained by:  (i) modelling each compact object by a point-mass, i.e.\
a Dirac-delta-function source, and (ii) regularizing the divergencies generated
by the use of delta-function sources by Riesz' analytic continuation method
\cite{riesz}.  It was also mentioned in \cite{D80} that the correct,
``effaced'' equations of motion could as well be obtained by regularizing the
2.5PN divergencies by using {\em dimensional regularization}.

More recently the 3PN equations of motion of two point-masses have been obtained
by two separate groups:  \cite{JS98,invariants,poincare} and \cite{BF1,BF4},
using different approaches, different coordinate gauges, and different
regularization methods.  The first group used the Arnowitt-Deser-Misner (ADM)
canonical approach \cite{ADM}, and derived the two-body Hamiltonian
corresponding to the ADM transverse-traceless (ADMTT) gauge.  It regularized
contact terms by Hadamard's partie finie, and divergent integrals by a
combination of a {\em Riesz-implemented Hadamard's partie finie} approach with
(when applicable) ordinary distribution theory \cite{JS98,J97,invariants}.  The
second group worked with the harmonically relaxed Einstein equations and derived
the equations of motion in a harmonic gauge.  It regularized the contact terms
and the divergent integrals by means of special-purpose variants of {\em
Hadamard's partie finie} approach applicable to a certain class of ``pseudo
functions'' \cite{BF2,BF3}.  In spite of the considerable efforts spent by both
groups, their regularization methods are defective in two (interconnected)
respects:  (i) they leave undetermined a dimensionless parameter ($\omega_s$, in
the notation of the first group) which was shown in \cite{lso} to be crucial for
determining the dynamics of the last few orbits, and (ii) they exhibit certain
mathematical inconsistencies.  By ``mathematical inconsistency'', we mean here
that both regularization methods defeat the very purpose for which they were
used, i.e.\ to uniquely define a perturbative solution of Einstein's theory
which depends only on two masses $m_1$, $m_2$.  We recall that Einstein's field
equations $E_{\mu \nu}\equiv R_{\mu \nu}-\frac{1}{2}\,R\,g_{\mu \nu} =
8\pi\,G\,T_{\mu\nu}$ exhibit the consistency feature that they imply the
equations of motion of the matter $(\nabla^{\nu}T_{\mu \nu}=0)$ as a
consequence of the (contracted) Bianchi identity
$\nabla^{\nu}E_{\mu\nu}\equiv0$.  To preserve this delicate consistency
property when perturbatively solving Einstein's equations in some gauge it is
necessary to use a regularization method which respects the basic properties of
the algebraic and differential calculus of ordinary functions:  such as the
associativity, commutativity and distributivity of point-wise addition and
multiplication, Leibniz's rule, Schwarz's rule
$(\pa_{\mu}\pa_{\nu}f=\pa_{\nu}\pa_{\mu}f)$, integration by parts, etc.
The regularization methods used up to now at the 3PN level violate at least one
of the basic properties of the standard calculus (for instance, Leibniz's rule
is violated both by the regularization method of \cite{JS98,invariants} and that
of \cite{BF2,BF3}).

We know only one regularization method which formally respects the basic
properties of standard calculus:  {\em dimensional regularization}
\cite{HV72,BG72,BM77,C84}.  It was invented as a way to preserve the
gauge-symmetry of perturbative quantum gauge theories.  Though our set up is
classical, our basic consistency problem is to respect the gauge-symmetry
(underlying the link between Bianchi identities and equations of motion) of
perturbative general relativity.  Dimensional regularization seems to be the
ideal tool for generating a consistent perturbative solution of Einstein's
theory depending only on $m_1$, $m_2$.  In this letter, we shall show how to use
dimensional regularization at the 3PN level in the ADM framework \cite{DJS}.
Our three main results will be:  (i) all the pole parts $\propto(d-3)^{-1}$
(where $d$ denotes the continued space dimension) appearing in intermediate
contributions cancel in the total 3PN Hamiltonian $H$, (ii) the contributions
to $H$ which are quadratic in momenta are unambiguously determined by
dimensional regularization to take the unique value $(\omega_k=41/24)$ which
preserves global Poincar\'e invariance \cite{poincare,BF1}, and (iii) the
momentum-independent contributions to $H$ are unambiguously determined and yield
the following unique value for the heretofore undetermined dimensionless
parameter entering 3PN dynamics:
\be
\label{eq1.1}
\omega_s = 0 \,.
\ee

\section{Dimensional continuation of the 3PN ADM Hamiltonian}\label{sec2}

Let $D\equiv{d+1}$ denote the (analytically continued) space-time dimension.
The ADM approach \cite{ADM} uses a $d+1$ split of the coupled gravity-matter
dynamics and works with the canonical pairs $(x_a^i,p_{ai})$ and 
$(g_{ij},\pi^{ij})$ ($i,j,k,\ldots$ denote spatial indices taking (formally) $d$ 
values; $a=1$, \ldots, $N$ labels the particles).  The dimensionally continued 
hamiltonian and momentum constraints read (in units where $16\pi\,G_D=1$)
\begin{mathletters}
\label{eq2.3}
\bea
\label{eq2.3a}
\sqrt{g}\,R &=& \frac{1}{\sqrt g} \left(g_{ik} \, g_{j\ell} \, \pi^{ij} \, 
\pi^{k\ell} - \frac{1}{d-1} \ (g_{ij} \, \pi^{ij})^2 \right)
+ \sum_a (m_a^2 + g_a^{ij} \, p_{ai} \, p_{aj})^{\frac{1}{2}} \, \delta_a \,,
\\[2ex]
\label{eq2.3b}
-2\,D_j\,\pi^{ij} &=& \sum_a g_a^{ij} \, p_{aj} \, \delta_a \,. 
\eea
\end{mathletters}
Here $\delta_a\equiv\delta({\bf x}-{\bf x}_a)$
(with $\int{d^dx\,\delta({\bf x})}=1$), 
$g_a^{ij}\equiv{g^{ij}({\bf x}_a)}$ (which will be seen to be perturbatively 
unambiguously defined and finite), and $D_j$ denotes the $d$-dimensional 
covariant derivative (acting on a tensor density).

The application of the ADM formalism to the gravitational dynamics of $N$ 
particles consists of four steps. Step (i) consists in fixing the gauge by 
demanding that $g_{ij}$ and $\pi^{ij}$ have the forms (ADMTT gauge)
\be
\label{eq2.5}
g_{ij} = A (\phi) \, \delta_{ij} + h_{ij}^{\rm TT} \,, \quad
\pi^{ij} = \widetilde{\pi}^{ij} (V^k) + \pi_{\rm TT}^{ij} \,,
\ee
where
\begin{mathletters}
\label{eq2.6}
\bea
\label{eq2.6a}
A(\phi) &\equiv& \left(1+\frac{d-2}{4(d-1)}\phi\right)^{4/(d-2)} \,,
\\[2ex]
\label{eq2.6b}
\widetilde{\pi}^{ij}(V^k) &\equiv& \pa_i V^j + \pa_j V^i
- \frac{2}{d}\, \delta^{ij}\, \pa_k V^k \,,
\eea
\end{mathletters}
and where the TT pieces $h_{ij}^{\rm TT}$, $\pi_{\rm TT}^{ij}$ are transverse 
and traceless, i.e.\ satisfy
$\pa_j f_{ij}^{\rm TT} = 0 = \delta^{ij}\,f_{ij}^{\rm TT}$ with $f=h$ or $\pi$.

Step (ii) consists in solving, by a perturbative expansion, the constraints
(\ref{eq2.3}) by expressing the ``longitudinal'' variables $\phi$ and
$\widetilde{\pi}^{ij}$ (i.e.\ $V^i$) in terms of the dynamical degrees of
freedom $(x_a^i,p_{ai};h_{ij}^{\rm TT},\pi_{\rm TT}^{ij})$.  This is done
by means of a post-Newtonian (PN) expansion:
$\phi = \phi_{(2)} + \phi_{(4)} + \phi_{(6)} + \phi_{(8)} + \phi_{(10)}$,
$V^i = V_{(3)}^i + V_{(5)}^i + V_{(7)}^i$.
Here, the numbers within parentheses denote the formal orders in the inverse 
velocity of light.  E.g., $\phi_{(2)}\sim{Gm/(c^2\,r^{d-2})}$. [Apart from this 
formal use of $1/c$ we set everywhere $c=1$.]  For instance, we have
\begin{mathletters}
\bea
\label{eq2.7}
\Delta\,\phi_{(2)} &=& - \sum_a m_a \, \delta_a \, ,
\\[2ex]
\label{eq2.8}
\Delta\,\phi_{(4)} &=& -\frac{1}{2} \sum_a \frac{{\bf p}_a^2}{m_a}\delta_a
+ \frac{d-2}{4(d-1)} \fii \sum_a m_a\delta_a \, ,
\\[2ex]
\label{eq2.9}
\pa_j \, \widetilde{\pi}_{(3)}^{ij} (V) &\equiv&
\Delta \, V_{(3)}^i + \left( 1 - \frac{2}{d} \right) \,\pa_{ij} V_{(3)}^j
= - \frac{1}{2} \sum_a p_{ai} \, 
\delta_a \, .
\eea
\end{mathletters}
The solution of Eq.\ (\ref{eq2.9}) can be written as
\be
\label{eq2.10}
V_{(3)}^i =  \frac{d-2}{4(d-1)} \sum_a p_{aj} 
\left(\Delta^{-2}\delta_a\right)_{,ij}
- \frac{1}{2} \sum_a p_{ai} \left(\Delta^{-1}\delta_a\right)\, .
\ee

Step (iii) consists in expressing the total energy of the system $E\,[\phi]$
(given by the surface integral at spatial infinity of $-\pa_i\,\phi$) in
terms of the (particle and gravitational) dynamical degrees of freedom.  This
yields an Hamiltonian $H(x_a^i,p_{ai},h_{ij}^{\rm TT},\pi_{\rm TT}^{ij})$
which describes the coupled dynamics of matter and gravity.  The fourth and
final step used in the ADM approach to the problem of motion of $N$ point masses
\cite{OOKH74,S85,DS85,JS98} consists in:  (iv) eliminating the gravitational
variables $(h_{ij}^{\rm TT},\pi_{\rm TT}^{ij})$ by perturbatively solving
their field equations (as obtained from varying the Hamiltonian derived in the
previous step).  This is again done by replacing $h_{ij}^{\rm TT}$ and $\pi_{\rm
TT}^{ij}$ by the PN-expanded solutions of their field equations.  [We are
interested here in the conservative dynamics and use a time-symmetric
(half-retarded half-advanced) Green function.  See \cite{S85,JS97} for the
discussion of radiation-reaction effects in the ADM approach, and
\cite{DS85,JS98,invariants} for the subtlety of the reduction of the higher
time-derivatives appearing when solving $(h_{ij}^{\rm TT} , \pi_{\rm TT}^{ij})$
in terms of the matter.]  These PN expansions start as:  $h_{ij}^{\rm TT} =
h_{(4)ij}^{\rm TT} + h_{(6)ij}^{\rm TT} + \cdots\,$, $\pi_{\rm TT}^{ij} =
\pi_{(5){\rm TT}}^{ij} + \pi_{(7){\rm TT}}^{ij} + \cdots\,$.  Actually, the fact
that the field equations for $h_{ij}^{\rm TT}$, $\pi_{\rm TT}^{ij}$ must be
satisfied implies that it is sufficient (at 3PN) to replace $(h_{ij}^{\rm TT} ,
\pi_{\rm TT}^{ij})$ by their leading contributions $(h_{(4)ij}^{\rm TT},
\pi_{(5){\rm TT}}^{ij})$.  These quantities satisfy the equations
\begin{mathletters}
\label{eq2.12}
\bea
\label{eq2.12a}
\Delta h_{(4)ij}^{\rm TT} &=&  \left( -\sum_a\frac{p_{ai}p_{aj}}{m_a}\delta_a
- \frac{d-2}{2(d-1)}{\gfii i}{\gfii j} \right)^{\text{TT}}\, ,
\\[2ex]
\label{eq2.12b}
\pi_{(5){\rm TT}}^{ij} &=& \frac{1}{2} {\dothttiv ij}
+ \frac{d-2}{d-1} (\fii{\pitiii ij})^{\text{TT}}\, .
\eea
\end{mathletters}
The superscripts TT in Eqs.\ (\ref{eq2.12}) denote the application to a second 
rank tensor of the $d$-dimensional (spatially nonlocal) TT-projection operator.

The matter Hamiltonian obtained after elimination of the gravitational variables
can be simplified by using many integration by parts allowing one to
successively eliminate $\phi_{(8)}$, $\phi_{(6)}$, $V_{(7)}^i$ and $V_{(5)}^i$
by using the elliptic equations they must satisfy.  [Some details of this
elimination process will be discussed in \cite{DJS}.]  For convenience, we have
performed this elimination procedure in a form closely parallel to the one
performed (when $d=3$) in \cite{invariants}.  Finally, the dimensional
continuation of the matter-reduced dimensionally continued 3PN Hamiltonian reads
(we do not write the $d$-dimensional versions of the non-problematic 1PN and 2PN
Hamiltonians)
\be
\label{eq2.15}
H_{\rm 3PN} =  -\frac{5}{128}\sum_a\frac{({\bf p}_a^2)^4}{m_a^7}
+ \int d^d\!x\,(h_1 + h_2 + h_3)\, ,
\ee
where
\begin{mathletters}
\label{eq2.16}
\bea
\label{eq2.16a}
h_1 &=& \left( \frac{(4-d)(d-2)^2}{64(d-1)^3}\siv\fii^2
- \frac{(4-d)(d-2)}{16(d-1)^2}\siv\fiv \right) \sum_a m_a\delta_a
\nonumber\\[2ex]&&
+ \left( -\frac{d(d+2)(3d-4)}{192(d-1)^3}\fii^3
+ \frac{(d+2)(3d-4)}{16(d-1)^2}\fii\fiv
+ \frac{(d-4)(d+2)}{32(d-1)^2}\siv\fii \right)
\sum_a \frac{{\bf p}_a^2}{m_a}\delta_a
\nonumber\\[2ex]&&
+ \left( -\frac{d(d+6)}{64(d-1)^2}\fii^2 + \frac{d}{8(d-1)}\fiv
-\frac{4-d}{32(d-1)}\siv \right) \sum_a \frac{({\bf p}_a^2)^2}{m_a^3}\delta_a
-\frac{d+4}{32(d-1)}\fii \sum_a \frac{({\bf p}_a^2)^3}{m_a^5}\delta_a
\nonumber\\[2ex]&&
+ \frac{4-d}{4(d-1)}\siv({\pitiii ij})^2
- \frac{3d-4}{2(d-1)}\fiv({\pitiii ij})^2
\nonumber\\[2ex]&&
+ \frac{(4-d)(d-2)}{4(d-1)^2}{\gfii i}{\gsiv j}{\httiv ij}
- \frac{(d-2)(2d-3)}{2(d-1)^2}{\gfii i}{\gfiv j}{\httiv ij}
+ \frac{1}{4} {\httiv ij} \sum_a \frac{{\bf p}_a^2 p_{ai} p_{aj}}{m_a^3}\delta_a
\nonumber\\[2ex]&&
- \left((\fii{\pitiii ij})^{\text{TT}}\right)^2
- \frac{d-2}{d-1} (\fii{\pitiii ij})^{\text{TT}}{\dothttiv ij}
- \frac{1}{4} \left({\dothttiv ij}\right)^2 \,,
\\[2ex]
\label{eq2.16b}
h_2 &=& \left( \frac{(d-2)^4}{128(d-1)^4}\fii^4
- \frac{(d-2)^3}{8(d-1)^3}\fii^2\fiv
+ \frac{(d-2)^2}{4(d-1)^2}\fiv^2 \right) \sum_a m_a\delta_a \,,
\\[2ex]
\label{eq2.16c}
h_3 &=& \frac{(3d-4)(3d-2)}{16(d-1)^2}\fii^2({\pitiii ij})^2
+ 2 \left({\gv ki}{\gv  kj}-{\gv ik}{\gv jk}
           -\frac{4}{d}{\gv ij}{\gv kk}\right) {\httiv ij}
\nonumber\\[2ex]&&
+ 4 {\v i}{\gv kj}{\ghttiv ijk}
+ \frac{(d-2)(3d-4)}{8(d-1)^3}\fii{\gfii i}{\gfii j}{\httiv ij}
\nonumber\\[2ex]&&
- \frac{d+2}{4(d-1)} \left(\fii{\httiv ij}\right)_{,k}{\gsij ijk}
- \frac{1}{2(d-1)} \left(\fii{\httiv ij}\right)_{,k}{\ghttiv ijk} \,.
\eea
\end{mathletters}

The 3PN Hamiltonian (\ref{eq2.15}) is expressed in terms of the following 
quantities: $\phi_{(2)}$ defined by Eq.\ (\ref{eq2.7}), $\phi_{(4)}$ defined by 
Eq.\ (\ref{eq2.8}), $V_{(3)}^i$ defined by Eq.\ (\ref{eq2.10}), 
$\widetilde{\pi}_{(3)}^{ij}\equiv\widetilde{\pi}^{ij}(V_{(3)}^k)$ defined by 
Eq.\ (\ref{eq2.6b}), $h_{(4)ij}^{\rm TT}$ defined by Eq.\ (\ref{eq2.12a}), as 
well as $S_{(4)}$ and $S_{(4)ij}$ defined by the following equations
\be
\label{eq2.17}
\siv \equiv \Delta^{-1} \sum_a \frac{{\bf p}_a^2}{m_a} \delta_a \,, \quad
{\sij ij} \equiv \Delta^{-1} \sum_a \frac{p_{ai}p_{aj}}{m_a} \delta_a \,.
\ee

\begin{figure}
\centerline{\epsfig{file=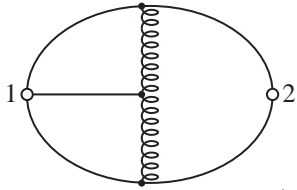}}
\caption{\label{Fig1}
``Three-loop diagram'' representing the contribution (of order 
$G^4\,m_1^3\,m_2^2$) proportional to
$\int{d^dx\,\phi_1\,\pa_k h_{(40)ij}^{\rm TT}\,\pa_k h_{(40)ij}^{\rm TT}}$ in 
the Hamiltonian, where $h_{(40)ij}^{\rm TT}\equiv
{-2\,c(d)\,\Delta^{-1}(\pa_i\phi_1\,\pa_j\phi_2)^{\rm TT}}$. The solid lines 
represent the ``propagator'' $\Delta^{-1}$ of $\phi$, the helicoidal lines 
represent the ``propagator'' $(\Delta^{-1})^{\rm TT}$ of $h_{(4)ij}^{\rm TT}$,
and the circles labelled 1 and 2 denote the sources $m_1\,\delta_1$ and
$m_2\,\delta_2$.}
\end{figure}

Using the natural (propagator)$*$(source) structure of each building block of
the Hamiltonian $H$, we can represent $H$ by a sum of diagrams.  For instance
the last term in Eq.\ (\ref{eq2.16c}) gives rise (after expanding it in powers
of $m_1$ and $m_2$) to several diagrams, one of which ($\propto m_1^3\,m_2^2$)
is represented (for illustration) in Fig.\ \ref{Fig1}.  Note that this diagram
contains three (classical) loops.  More generally the most nonlinear
(momentum-independent) contributions to $H_{n{\rm PN}}$ contain $n$ loops.

\section{Finiteness of $H_{\rm ADM}(\lowercase{d})$ as $\lowercase{d}\to3$} 
\label{sec3}

The perturbative consistency of dimensional regularization \cite{C84} ensures
that the formal $d$-dimensional PN-expanded solution $g_{\mu\nu}(x^{\lambda})$
given in Section~\ref{sec2} will directly generate, when $d$ is analytically
continued up to $d=3$, a 3-dimensional solution of Einstein's theory {\em if} it
is finite as $d\to3$, i.e.\ if no poles proportional to $(d-3)^{-1}$
arise when $d\to3$.  One can analytically find out all the potentially
``dangerous'' terms (those which might give rise to a pole) in $H_{\rm 3PN}$,
Eq.\ (\ref{eq2.15}), by the following analysis.  First, the $d$-dimensional
solution of the basic equation $\Delta{u_a}=-\delta_a$ is 
${u_a}\equiv{-\Delta^{-1}\delta_a}=k\,r_a^{2-d}$, where
$k\equiv{\Gamma\left(\frac{d-2}{2}\right)/(4\pi^{\frac{d}{2}})}$ and
${r_a}\equiv{\vert{\bf x}-{\bf x}_a\vert}$. [Though $k$ depends on $d$ we shall 
factor the various powers of $k$ entering each contribution to $H_{\rm 3PN}$ 
without ever expanding its dependence on $d$ as $d\to3$.]  This allows one to 
write down explicitly $\phi_{(2)}$, $\phi_{(4)}$, $S_{(4)}$, and $S_{(4)ij}$.  
The explicit expression of $V_{(3)}^i$ is then obtained by iterating the Poisson 
operator $\Delta^{-1}$ using 
$\Delta^{-1}r_a^{\lambda}=r_a^{\lambda+2}/\left((\lambda+2)(\lambda+d)\right)$.

These results suffice to write down the part of $h_{(4)ij}^{\rm TT}$, Eq.\ 
(\ref{eq2.12a}), whose source is $-m_a^{-1}\,p_{ai}\,p_{aj}\,\delta_a$.  We 
denote this part as $h_{(42)aij}^{\rm TT}$ (with $a = 1,2$).  The
part of $h_{(4)ij}^{\rm TT}$ which does not depend on momenta will be denoted by
$h_{(40)ij}^{\rm TT}$.  One can control its crucial ``ultra violet'' (UV)
behaviour as $r_a\to0$ (which suffices to analyze the potentially
dangerous terms) in the following way.  First, let $c(d)\equiv(d-2)/(2(d-1))$ so 
that the source term for $h_{(40)ij}^{\rm TT}$ is the TT projection
of $-c(d)\,\sigma_{ij}(\phi_{(2)},\phi_{(2)})$ where
$\sigma_{ij}(\phi,\psi)\equiv{\pa_i\phi\,\pa_j\psi}$.  Inserting 
the solution $\phi_{(2)}=m_1\,u_1+m_2\,u_2$ into $\sigma_{ij}$, and applying the 
TT operator (which projects to zero $\sigma_{ij}(u_1,u_1)$ and 
$\sigma_{ij}(u_2,u_2)$), yields
$h_{(40)ij}^{\rm TT}=-2\,c(d)\,m_1\,m_2\,
\Delta^{-1}(\sigma_{ij}(u_1, u_2))^{\rm TT}$.
Operating by parts under the TT operator further yields
$h_{(40)ij}^{\rm TT}=+2\,c(d)\,m_1\,m_2\,
\Delta^{-1}(u_1\,\pa_{ij}u_2)^{\rm TT}$. It is then easily proven that, near 
${\bf x}_1$, $h_{(40)ij}^{\rm TT}$ is the sum of a regular $(C^{\infty})$ 
function and of a singular expansion in powers of $r_1$ obtained by inserting in 
the source term $\propto(u_1\,\pa_{ij}u_2)^{\rm TT}$ the Taylor expansion of
$\pa_{ij}u_2({\bf x})$ near ${\bf x}={\bf x}_1$, and applying the operator 
$\Delta^{-1}$ to each term of the expansion.

Using this local UV analysis we studied the convergence, near, say, ${\bf x}_1$,
of the integral giving $H_{\rm ADM}$.  The conclusion is that poles $\propto
(d-3)^{-1}$ can arise only from the third integrand $h_3$, Eq.\ (\ref{eq2.16c}).
More precisely, one finds that, when expanding each field entering $h_3$ in
powers of the masses $m_1$ and $m_2$, and then expanding each partial integrand
in powers of $r_1$ there arise {\em ten} (and only ten) separate terms $T_A$,
$A=1$, \ldots, 10, giving rise to poles as $d\to3$.  These pole-generating 
partial integrands have all the structure (for the poles generated by the 
integration near $r_1=0$)
\bea
\label{eq3.3}
T_A &=& k^4 \, m_1 \, m_2 \, r_1^{6-3d} \, \Big(
c_{A1} \, D(p_1,p_1) + c_{A2} \, (n_1\cdot p_1) \, D(n_1,p_1)
\nonumber\\[2ex]&&
+ \, c_{A3} \, (n_1\cdot p_1)^2 \, D(n_1,n_1)
+ c_{A4} \, p_1^2 \, D(n_1,n_1) \Big) \,,
\eea
where $c_{A1}$, \ldots, $c_{A4}$ are $d$-dependent coefficients and where
$D_{ij}\equiv\left[\pa_{ij}r_2^{2-d}\right]_{{\bf x}={\bf x}_1}$,
$D(p,q)\equiv D_{ij}\,p^i\,q^j$,
$n_1^i\equiv r_1^{-1} (x^i - x_1^i)$, and
$n_1\cdot p_1\equiv n_1^i\,p_{1i}\equiv\delta_{ij}\,n_1^i\,p_1^j$.  These ten 
terms arise from the terms in the mass-expanded version of $h_3$ with the 
respective structures:
$\phi_1 \,\phi_2 \, (\pa V_1)^2$,
$\pa V_1 \, \pa V_1 \,h_{(40)}^{\rm TT}$,
$V_1 \, \pa V_1 \, \pa h_{(40)}^{\rm TT}$,
$\phi_2 \, \pa\phi_1 \, \pa\phi_1 \, h_{(42)1}^{\rm TT}$,
$\phi_1 \, \pa\phi_2 \, \pa\phi_1 \, h_{(42)1}^{\rm TT}$,
$\phi_1 \, \pa h_{(42)1}^{\rm TT} \, \pa h_{(40)}^{\rm TT}$,
$\pa S_1 \, \pa\phi_1 \, h_{(40)}^{\rm TT}$,
$\pa S_1\, \phi_1 \, \pa h_{(40)}^{\rm TT}$,
$\pa\phi_1 \, \pa h_{(42)1}^{\rm TT} \, h_{(40)}^{\rm TT}$, and
$\pa\phi_1 \,h_{(42)1}^{\rm TT} \, \pa h_{(40)}^{\rm TT}$.
Here we have suppressed, for readibility, the PN indices $\scriptstyle (2)$ (on
$\phi$), $\scriptstyle (3)$ (on $V$) and $\scriptstyle (4)$ (on $S$), but we
left the indices $\scriptstyle (40)$ and $\scriptstyle (42)$ on $h^{\rm TT}$.
The explicit labels 1 or 2 refer to their sources $\propto\delta_1$ or
$\delta_2$.  Each term $T_A$ carries its full coefficient, as arising from the
successive (mass-expanded and Taylor-expanded) terms in $h_3$, Eq.\
(\ref{eq2.16c}).

One then computes the ``local'' contribution (in a ball around ${\bf x}_1$ of 
radius $\ell_1$, with $0<{\ell_1}\ll{r_{12}\equiv
\vert{\bf x}_1-{\bf x}_2\vert}$) of each ``dangerous'' term $T_A$.  After 
averaging (in a $d$-dimensional sense) over angles, and integrating over the 
radius $r_1$, we find that each ``dangerous'' term locally yields, when $d \to 
3$, the following ``pole''contribution to the Hamiltonian~(\ref{eq2.15})
\be
\label{eq3.8}
H_A^{\rm loc} (d) = - \frac{1}{2} \ \Omega_d \, k^4 \, m_1 \, m_2 \, D(p_1,p_1) 
\, \ell_1^{-2(d-3)} \, \frac{c_A (d)}{d-3} \, ,
\ee
where $\Omega_d$ is the area of the unit sphere in $d$ dimensions, and where we 
defined
\be
\label{eq3.7}
c_A(d) \equiv c_{A1}(d) + \frac{c_{A2}(d)}{d} + \frac{2\,c_{A3}(d)}{d(d+2)} \,.
\ee

We have explicitly computed the $10\times4=40$ $d$-dependent coefficients
$c_{A1}(d)$, \ldots, $c_{A4}(d)$ as well as the 10 angular-averaged
coefficients $c_A(d)$.  Note that the value of each of these coefficients
delicately depends on having correctly derived the $d$-dependent coefficients
appearing in the Hamiltonian, as well as all the $d$-dependent factors arising
from the $\Delta^{-1}$ and TT operators and from the angular averaging.  The
results are listed in Table~I, each coefficient being expanded in powers of
$\ve\equiv{d-3}$, e.g.\ 
$c_A (d)=c_A (3)+\ve\,c'_A (3)+{\cal O}(\ve^2)$. A first important result from 
Table~I is that, as one checks, the sum of the ten 3-dimensional averaged 
coefficients $c_A(3)$ exactly vanishes:
\be
\label{eq3.9}
\sum_{A=1}^{10} c_A (3) = 0 \,.
\ee
In view of Eq.\ (\ref{eq3.8}) this proves that the total pole part, as $d\to3$, 
of $H(d)$ vanishes.  In other words, dimensional continuation up to $d=3$ gives 
a unique, finite value for $H_{\rm 3PN}^{\rm ADM}$.  [This result confirms the 
finding of \cite{JS98,invariants} that all the ``Riesz poles'' appearing in the 
Riesz-based implementation of Hadamard's partie finie finally cancel.  However, 
this finding was not sufficient for being able to unambiguously compute the 
``regularized'' value of $H_{\rm 3PN}^{\rm ADM}$.]

\begin{table}[ht]
\caption{Expansion in powers of $\ve\equiv{d-3}$ of the $d$-dependent 
coefficients $c_{A1}(d)$, \ldots, $c_{A4}(d)$ and $c_A(d)$ entering the 10 
``dangerous'' terms $T_A$, $A=1$, \ldots, 10.}
\begin{center}
\begin{tabular}{cccccc}
$A$ & $c_{A1}$ & $c_{A2}$ & $c_{A3}$ & $c_{A4}$ & $c_{A}$ \\[2ex] \hline
&&&&&\\
1
& 0
& 0
& ${\dst \frac{315}{1024} + \frac{3711}{4096}\,\ve }$
& ${\dst \frac{315}{2048} + \frac{267}{1024}\,\ve  }$
& ${\dst \frac{21}{512}   + \frac{1013}{10240}\,\ve  }$
\\[3ex]
2
& ${\dst -\frac{13}{256} - \frac{263}{1536}\,\ve }$
& ${\dst -\frac{3}{128}  - \frac{47}{256}\,\ve }$
& ${\dst  \frac{15}{512} + \frac{33}{256}\,\ve }$
& ${\dst  \frac{5}{512}  + \frac{235}{3072}\,\ve }$
& ${\dst -\frac{7}{128}  - \frac{1649}{7680}\,\ve  }$
\\[3ex]
3
& ${\dst  \frac{7}{128} + \frac{487}{6144}\,\ve }$
& ${\dst -\frac{37}{256} - \frac{91}{384}\,\ve }$
& ${\dst -\frac{29}{1024} - \frac{665}{6144}\,\ve }$
& ${\dst  \frac{49}{1024} + \frac{49}{768}\,\ve }$
& ${\dst  \frac{7}{2560}  + \frac{601}{153600}\,\ve }$
\\[3ex]
4
& 0
& 0
& ${\dst  \frac{15}{512} + \frac{191}{2048}\,\ve }$
& ${\dst -\frac{5}{512}  - \frac{57}{2048}\,\ve }$
& ${\dst  \frac{1}{256}  + \frac{53}{5120}\,\ve }$
\\[3ex]
5
& 0
& ${\dst -\frac{5}{64}  - \frac{37}{256}\,\ve }$
& ${\dst -\frac{5}{128} - \frac{57}{512}\,\ve }$
& ${\dst  \frac{5}{128} + \frac{37}{512}\,\ve }$
& ${\dst -\frac{1}{32}  - \frac{33}{640}\,\ve }$
\\[3ex]
6
& ${\dst  \frac{43}{1536} + \frac{7}{256}\,\ve }$
& ${\dst  \frac{31}{768}  + \frac{221}{1536}\,\ve }$
& ${\dst -\frac{73}{3072} - \frac{61}{6144}\,\ve }$
& ${\dst -\frac{17}{3072} - \frac{263}{6144}\,\ve }$
& ${\dst  \frac{49}{1280} + \frac{5467}{76800}\,\ve }$
\\[3ex]
7
& ${\dst  \frac{65}{768}  + \frac{421}{1536}\,\ve }$
& ${\dst -\frac{25}{384}  - \frac{35}{256}\,\ve }$
& ${\dst -\frac{5}{1536}  - \frac{37}{3072}\,\ve }$
& ${\dst  \frac{35}{1536} + \frac{43}{1024}\,\ve }$
& ${\dst  \frac{1}{16}    + \frac{15}{64}\,\ve }$
\\[3ex]
8
& ${\dst -\frac{65}{768}  - \frac{161}{1536}\,\ve }$
& ${\dst  \frac{25}{384}  + \frac{5}{768}\,\ve }$
& ${\dst  \frac{5}{1536}  + \frac{17}{3072}\,\ve }$
& ${\dst -\frac{35}{1536} + \frac{11}{3072}\,\ve }$
& ${\dst -\frac{1}{16}    - \frac{7}{64}\,\ve }$
\\[3ex]
9
& ${\dst -\frac{13}{3072} - \frac{23}{1536}\,\ve }$
& ${\dst -\frac{19}{1536} - \frac{49}{768}\,\ve }$
& ${\dst  \frac{7}{6144}  - \frac{11}{4096}\,\ve }$
& ${\dst  \frac{23}{6144} + \frac{257}{12288}\,\ve }$
& ${\dst -\frac{21}{2560} - \frac{5423}{153600}\,\ve }$
\\[3ex]
10
& ${\dst  \frac{13}{3072} + \frac{5}{768}\,\ve }$
& ${\dst  \frac{19}{1536} + \frac{5}{128}\,\ve }$
& ${\dst -\frac{7}{6144}  + \frac{61}{12288}\,\ve }$
& ${\dst -\frac{23}{6144} - \frac{55}{4096}\,\ve }$
& ${\dst  \frac{21}{2560} + \frac{2903}{153600}\,\ve }$
\\&&&&& \\
\end{tabular}
\end{center}
\end{table}

\section{Dimensional-continuation determination of the Hamiltonian}\label{sec4}

Having shown that $H_{\rm 3PN}^{\rm ADM}(d)$ has a unique, finite value as 
$d\to3$, it remains the crucial task of explicitly computing this finite value.  
Let us first focus on the $h_3$-generated part of the Hamiltonian, i.e.\ on 
$H_3(d)\equiv\int{d^dx\,h_3(d)}$, with $h_3(d)$ given by Eq.\ (\ref{eq2.16c}), 
and on the finite limit:  $\lim_{d\to3}\,H_3(d)$.  Let us compute the difference 
$\Delta{H_3}\equiv\lim_{d\to3}\,H_3 (d)-H_3^{\rm DJS}$, where
$H_3^{\rm DJS}\equiv{\rm RH}\left[\int{d^3x\,h_3^{\rm DJS}}\right]$ is the 
Riesz-implemented Hadamard regularization (RH), as defined in 
\cite{JS98,J97,invariants}, of the 3-dimensional integral of the integrand 
$h_3^{\rm DJS}$ given by Eq.\ (A9c) of \cite{invariants}.  Note that {\em each 
term} in $h_3(d)$, Eq.\ (\ref{eq2.16c}), tends, when $d\to3$, to a corresponding 
term (with the same coefficient in the limit) in $h_3^{\rm DJS}$.  The 
comparison between the two integrals $H_3 (d)$ and $H_3^{\rm DJS}$ is best done 
by:  (i) separating the full space integral into
two {\em local} integrals (restricted to balls of radii $\ell_1$ and $\ell_2$
around ${\bf x}_1$ and ${\bf x}_2$) and a global
integral over the rest of space, and (ii) decomposing each term of each
integrand into a {\em regular part} (which is absolutely convergent in $d=3$) 
and a {\em singular part} (which is not absolutely convergent in $d=3$) made of 
a finite number of terms of the form (say, near ${\bf x}_1$) 
$r_1^{-\lambda}\,C_{i_1\cdots i_k}\,n_1^{i_1}\cdots n_1^{i_k}$.  It is then seen 
that the difference $\Delta{H_3}$ is simply
given by the difference between $\lim_{d\to3}\,H_3^{{\rm loc}\,{\rm sing}}(d)$ 
and ${\rm RH}\,[H_{\rm 3DJS}^{{\rm loc}\,{\rm sing}}]$,
involving only the {\em local} ($r_1<\ell_1$ or $r_2<\ell_2$) integrals of
the {\em singular} parts of the integrands.  Next, one checks that all local
singularities which do not correspond to poles in $d-3$ are given, in the limit
$d \to 3$, the same regularized values in both regularization methods.
However, the situation is different for the local (``logarithmic'')
singularities $\propto{r_1^{-3}}$ in $d=3$, which give rise to poles in $d\ne3$. 
 The complete list of these ``dangerous'' local singularities has been given
above:  they are the ten terms $T_A$ listed in Table~I.  [Note that they are all
quadratic in $p_1$.  Here, we focus on the UV behaviour near
${\bf x}_1$; one should add the similar terms obtained by exchanging
$1\leftrightarrow2$.]  After averaging over angles, the RH regularization of
the $d=3$ limit of each local singular integral is $\propto 
c_A(3)\ln(\ell_1/s_1)$ where $s_1$ is the scale entering the definition of the 
RH operation.
After summing over $A=1$, \ldots, $10$, the result (\ref{eq3.9}) shows that
${\rm RH}\,(H_{\rm 3DJS}^{{\rm loc}\,{\rm sing}})=0$.  On the other hand,
dimensional continuation gives for the value of $H_3^{{\rm loc}\,{\rm sing}}(d)$ 
the sum over $A$ of the values (\ref{eq3.8}).  Finally, we conclude, using
$c_A(3+\ve)=c_A(3)+\ve\,c'_A(3)+{\cal O}(\varepsilon^2)$,
Eq.\ (\ref{eq3.9}), and the 3-dimensional limits $\Omega_3=4\pi$, 
$k(d=3)=1/(4\pi)$, that ($r_{12}\equiv\vert{\bf x}_1-{\bf x}_2 \vert$)
\be
\label{eq4.1}
\Delta{H_3} \equiv \lim_{d\to3} H_3(d) - H_3^{\rm DJS}
= -\frac{1}{2} \left(  \sum_{A=1}^{10} c'_A (3) \right)
\frac{m_1\,m_2}{(4\pi)^3} \ p_{1i} \, p_{1j} \, \pa_{ij}\frac{1}{r_{12}}
+ 1\leftrightarrow2 \,.
\ee

We have similarly evaluated
$\Delta{H_i}\equiv\lim_{d\to3}\,H_i(d)-H_i^{\rm DJS}$, for $i=1,2$, where 
$H_i(d)\equiv\int{d^dx\,h_i(d)}$ and
$H_i^{\rm DJS}\equiv{\rm RH}\left[\int{d^3x\,h_i^{\rm DJS}}\right]$.  Here
the formal expressions of $h_1(d)$ and $h_2(d)$ do not seem to correspond to
the expressions of $h_1^{\rm DJS}$, $h_2^{\rm DJS}$ given in Eqs.\ (A9a), (A9b) 
of \cite{invariants}.  However, we can use the properties of dimensional
regularization to transform $h_1(d)$ and $h_2(d)$.  We have checked that, by
using suitable integration by parts (which are allowed in dimensional
continuation), one could transform $h_i(d)$, $i=1,2$, into new expressions
$h'_i(d)$ which are term-by-term similar to $h_i^{\rm DJS}$, in the same
sense that $h_3(d)$ was similar to $h_3^{\rm DJS}$.  An analysis (using the
tools given above) of the UV behaviour of $h'_i(d)$ then shows that they
contain no ``dangerous'' terms ${\cal O}(r_1^{-\lambda (d)})$ with 
$\lambda(3)=3$.  The reasoning of the previous section then shows that
$\Delta{H_1}=\Delta{H_2}=0$.

Finally, we conclude that the finite value of the total Hamiltonian is given by: 
$\lim_{d\to3}H_{\rm 3PN}^{\rm ADM}(d)=H_{\rm 3PN}^{\rm DJS}+\Delta{H_3}$ with 
$\Delta{H_3}$ given by Eq.\ (\ref{eq4.1}).  The two main consequences of this 
result concern the two dimensionless parameters that were left undetermined by 
the original Riesz-Hadamard regularization of the 3PN Hamiltonian \cite{JS98}:  
$\omega_k$ and $\omega_s$.

First, from the definition (see, e.g., Eq.\ (13) of \cite{poincare}) of the
dimensionless ``kinetic'' parameter $\omega_k$ and the fact that $H_3^{\rm DJS}$
corresponds, by convention, to $\omega_k = 0$, the result (\ref{eq4.1}) is
equivalent to saying that dimensional continuation uniquely determines the value
of $\omega_k$ to be (using $(4\pi)^{-1} = 4G$)
\be
\label{eq4.2}
\omega_k^{{\rm dim} \, {\rm reg}} = 64 \sum_{A=1}^{10} c'_A (3) \,.
\ee
The sum of the $c'_A (3)$ read from Table~I is checked to be $\frac{41}{1536}$.
The dimensional continuation prediction (\ref{eq4.2}) therefore yields
$\omega_k^{{\rm dim}\,{\rm reg}}=\frac{41}{24}$.  As shown in
\cite{poincare} (see also \cite{BF1,BF4}) this value is the unique, correct
value which ensures that the regularized Hamiltonian admits a nonlinear
realization of global Poincar\'e invariance.  We find it truly remarkable that
the use of dimensional regularization in the ADM formalism (which violates from
the start manifest Poincar\'e invariance by splitting space and time and by
using a non Poincar\'e-invariant gauge) uniquely predicts, after a long
calculation involving many intermediate complicated $d$-dependent coefficients
and finally summing 50 different contributions of Table~I (indeed $c'_A (3)$ is
a combination of $c'_{A1}$, $c'_{A2}$, $c'_{A3}$, $c_{A2}$, and $c_{A3}$ for 
$A=1$, \ldots, 10), the unique, physically correct value of $\omega_k$.  We
interpret this as a very strong confirmation of the mathematical (and physical)
consistency of dimensional regularization, and, in particular, of its property
of perturbatively respecting gauge symmetry (i.e., in our case, diffeomorphism
invariance).

Second, as $\Delta{H_3}$ affects only the part of the Hamiltonian which is
quadratic in the momenta, we conclude that the heretofore undetermined ``static
parameter'' $\omega_s$ is simply:  $\omega_s=0$ (as it was in the ``reference''
Hamiltonian $H^{\rm DJS}$).  Note that this simple conclusion was
obtained after a detailed analysis of the UV behaviour of all the terms in
(\ref{eq2.15}).  In this analysis, the ADMTT gauge played a very useful role in
suppressing many of the stronger UV divergencies that occur in a harmonic gauge.
For instance the contribution $I_1(d)
=\int{d^dx\,\phi_1\,\pa_k h_{(40)ij}^{\rm TT}\,\pa_k h_{(40)ij}^{\rm TT}}$ where 
$h_{(40)ij}^{\rm TT}\equiv{-c(d)\,\Delta^{-1}(\pa_i\phi\,\pa_j\phi)^{\rm TT}}$, 
which corresponds to the diagram depicted in Fig.~\ref{Fig1}, turns out to be
completely ``non dangerous'' (and even locally integrable), while its
harmonic-gauge analog would contain
$U_{ij}\equiv{-c(d)\,\Delta^{-1}(\pa_i\phi\,\pa_j\phi)}$ instead of 
$h_{(40)ij}^{\rm TT}$ and would give rise
to several (gauge-spurious) ``dangerous'' integrals generating pole parts (and
logarithms in $d=3$).  [The existence of poles at 3PN, in harmonic coordinates,
was first mentioned in \cite{D83} and has been confirmed by the explicit
results of \cite{BF4}.]  Let us also note that dimensional regularization leads
to a very simple and consistent treatment of ``contact terms''.  Starting from
the best available definition of dimensional regularization, which is in the
Fourier domain (e.g.  \cite{C84}), we have checked that the value of, e.g.,
$I_2(d) = \int{d^dx\,\phi^4({\bf x})\,\delta_1}$ is $\phi^4({\bf x}_1)
= (m_2\,u_2({\bf x}_1))^4$.  This simple ``threading'' property of contact terms
extends to all the perturbative contact terms arising when formally solving
Einstein's equations.  Here, dimensional regularization differs from some
variants of Hadamard's partie finie regularization in which
${\rm Pf}\,(\phi^4)\ne\left({\rm Pf}\,(\phi)\right)^4$.

To conclude this letter let us mention some of the physical applications of our
results.  The most important is that our unique determination of the last
missing parameter in the 3PN Hamiltonian allows one to make full use of the 3PN
accuracy and, in particular, to analytically estimate the characteristics of the
last stable orbits (LSO) of binary black holes.  For instance, using the best
available PN-resummation method, namely the (Pad\'e-resummed) effective one-body
approach \cite{BD99,lso,D01}, one finds that the LSO binding energy of two
non-spinning equal-mass black holes is 1.67\%$(m_1+m_2)c^2$.  More generally,
our work allows one to extend to 3PN the approach proposed in \cite{BD99},
namely to start the numerical simulation of the coalescence of two black holes
(without or with spin \cite{D01}) just after they cross their LSO, so that they
are less than one orbit away from coalescence.

Our work contains another important lesson for numerical relativity:  our
dimensional-continuation calculation of $H_3$ and $\omega_k$ shows (as a
corollary) that the ``conformally flat'' truncation (i.e.\ setting
$h_{ij}^{\rm TT}$ to zero) which is of general use in numerical relativity, is
``inconsistent'' in that it ruins the delicate cancellation of poles and leads
to a formally infinite 3PN Hamiltonian.  [In addition, it violates global
Poincar\'e invariance.]  As the physical meaning of the 3PN pole cancellation is
that the perturbative algorithm unambiguously determines the unique, physically
correct solution of Einstein's equations depending on $m_1$ and $m_2$ which
matches two black holes \cite{DJS}, the lack of uniqueness associated to the
arbitrariness in subtracting the 3PN pole generated by the conformally flat
truncation probably explains why current initial data simulations
\cite{baumgarte,brug} significantly differ from the resummed analytical
estimates of LSO characteristics.

Finally, our work having convincingly established the perturbative consistency
of dimensional regularization, we strongly recommend to use it to uniquely
determine the 3PN contribution to the gravitational wave flux, whose
determination might otherwise be marred by spurious ambiguities.  Let us also
mention that it would be nice to independently confirm our results by two
different calculations:  (i) to use dimensional continuation, instead of a
variant of Hadamard regularization \cite{BF2,BF3}, to compute the 3PN dynamics
in harmonic coordinates, and (ii) to use dimensional continuation to recompute
$\omega_s$ by studying the motion of a test mass around a Schwarzschild black
hole to the first ``post-test-mass'' approximation (${\cal O}(\mu/M)$
corrections).

\section*{Acknowledgments}

This work was supported in part by the KBN Grant No.\ 5 P03B 034 20 (to P.J.).

\end{document}